%% file: hql06-lcerrito.tex
\documentclass[12pt]{article}
\usepackage{epsfig}

\input econfmacros.tex

\def\Title#1{\begin{center} {\Large {\bf #1} } \end{center}}

\begin{document}

\Title{Measurements of the Top Quark at the Tevatron Collider}

\begin{center}{\large \bf Contribution to the proceedings of HQL06,\\
Munich, October 16th-20th 2006}\end{center}

\bigskip\bigskip


\begin{raggedright}  

{\it Lucio Cerrito\index{Cerrito, L.}\\
on behalf of the CDF and D$\oslash$ Collaborations\\
Particle Physics Research Centre\\
Department of Physics\\
Queen Mary, University of London\\
E1 4NS London, U.K.}
\bigskip\bigskip
\end{raggedright}

\section{Introduction}
We present recent preliminary measurements of the top-antitop pair production cross section and determinations of the top quark pole mass, performed using the data collected by the CDF and D$\oslash$ Collaborations at the Tevatron Collider(\footnote{The Tevatron is a proton-antiproton synchrotron accelerator producing collisions at a centre-of-mass energy of 1.96 TeV (Run II) in two locations (CDF and D$\oslash$). The Tevatron operated until 1998 (Run I) at a centre-of-mass energy of 1.8 TeV.}). In the lepton plus jets final state, with semileptonic $B$ decay, the pair production cross section has now been measured at CDF using $\sim$760 pb$^{-1}$ of proton-antiproton collisions at a centre-of-mass energy of $\sqrt{s}$=1.96 TeV. A measurement of the production cross section has also been made with $\sim$1 fb$^{-1}$ of data in the all-jets final state by the CDF Collaboration. The mass of the top quark has now been measured using $\sim$1 fb$^{-1}$ of collision data using all decay channels of the top quark pair, yielding the most precise measurements of the top mass to date.

Top quarks are produced at the Tevatron Collider predominantly in pairs of top-antitop via quark-antiquark scattering ($\sim$85\%) and gluon-gluon fusion ($\sim$15\%). The decay of top quarks proceeds before hadronization almost exclusively as $t\rightarrow Wb$, therefore it is common to classify the final state of a $t\bar{t}$ event according to the decay modes of the $W$ bosons. The decay channels are labelled as: {\it dilepton}, when both $W$'s from the top pair decay leptonically, {\it lepton plus jets} when one the $W$'s decays leptonically and the second decays hadronically, and {\it all hadronic}, when both $W$'s decay hadronically. The identification of $b$ quarks in jets is made either through the measurement of a displaced secondary vertex in the event (due to the long lifetime of $B$-hadrons), or through the detection of a muon or electron from the semileptonic decay of $B$-hadrons. 

The measurement of the $t\bar{t}$ production cross section provides a test of the theory of Quantum Chromo Dynamics (QCD), while the comparison of the measurements from different decay channels allows probing both the production and decay mechanism described by the standard model (SM). At $\sqrt{s}$=1.96 TeV, the predicted $t\bar{t}$ cross section is $\sigma_{t\bar{t}}=6.8^{+0.7}_{-0.9}$ pb for a top mass ($M_t$) of 175 GeV/$c^2$~\cite{Cacciari}\cite{nikos}. The top quark mass, on the other hand, is a crucial ingredient of the SM and its precision measurement, combined with the independent measurement of the mass of the $W$ boson, allows predicting the SM Higgs boson mass.

\section{Top Quark Pair Production Cross Section}
\subsection{Lepton plus jets events with soft muon b-tagging}
This analysis is based on data corresponding to an integrated luminosity of $\sim$760 pb$^{-1}$, collected with the CDF II detector \cite{CDF,CDF2,CDF3,CDF4} between March 2002 and September 2005. The data are collected with an inclusive lepton trigger that required an electron (muon) with $E_T>$ 18 GeV ($P_T>$ 18 GeV/$c$)\cite{definitions}. From the inclusive lepton dataset we select events offline with a reconstructed isolated electron $E_T$ (muon $P_T$) greater than 20 GeV, missing $E_T>$ 20 GeV and at least 3 jets with $E_T>$ 15 GeV. Background events, predominantly due to QCD production of $W$ bosons with multiple jets, are rejected in the first instance using the total event energy $H_T>$ 200 GeV ($H_T$ is the scalar sum of the electron $E_T$, muon $P_T$, missing $E_T$ and jet $E_T$ for jets with $E_T>$ 8 GeV and $|\eta|<2.4$). To further enhance the top quark signal against background events, we identify events with one or more $b$-jets by searching inside jets for semileptonic decays of $B$-hadrons into muons (with $P_T>$ 3 GeV/$c$, $\Delta R<$ 0.6 from a jet axis). From 85 candidate $t\bar{t}$ events and 27.3$\pm$2.5 expected background events, and considering the detector acceptance and event selection efficiency, we determine a cross section of 7.8$\pm$1.7(stat)$^{+1.1}_{-1.0}$(syst) pb. The result is in good agreement with our previous measurement \cite{sltprd}, which used an integrated luminosity of $\sim$200 pb$^{-1}$ or about 1/4th of the presently analysed dataset.

\subsection{All hadronic events}\label{sec:2.2}
This analysis is performed using $\sim$1.0 fb$^{-1}$ of $p\bar{p}$ collisions at $\sqrt{s}$=1.96 TeV, selected with a multijet trigger. The trigger requires at least four calorimeter clusters with $E_T\ge$ 15 GeV and a total transverse energy exceeding 175 GeV. Candidate $t\bar{t}$ events are required to have between 6 and 8 jets with $E_T>$ 15 GeV and $|\eta|<$2.0, and jets must be separated in $\eta-\phi$ by $\Delta R\ge$0.5. Due to the low signal (S) to background (B) ratio after the trigger and topoligical selections defined above, a neural network (NN) based kinematical selection is applied. The NN is a Multilayer perceptron (MLP\cite{mlp}), a feed-forward network with an input layer, hidden layers and an output layer. The inputs to the NN include variables such as the total and sub-leading transverse energy, centrality, aplanarity and dijet and trijet masses \cite{prl79}. The single output node provides the variable $NN_{out}$ on which we base the selection. The final additional requirement is that at least one jet has a displaced secondary vertex (called a tag), compatible with the presence of one or more $b$-jets in the event. The best selection cut on $NN_{out}$ is determined considering the total expected signal and background and maximising $\Delta(B+S)/S$. We observe 1233 tags from within the selected sample, with a background expectation of 846$\pm$37 tags. The excess of tags with respect to the background is ascribed to the $t\bar{t}$ production and from this excess we measure a production cross section of: 8.3$\pm$1.0(stat)$\pm$0.5(lum)$^{+2.0}_{-1.5}$(syst) pb.

\begin{figure}[htb]
\begin{center}
\begin{tabular}{cc}
\epsfig{file=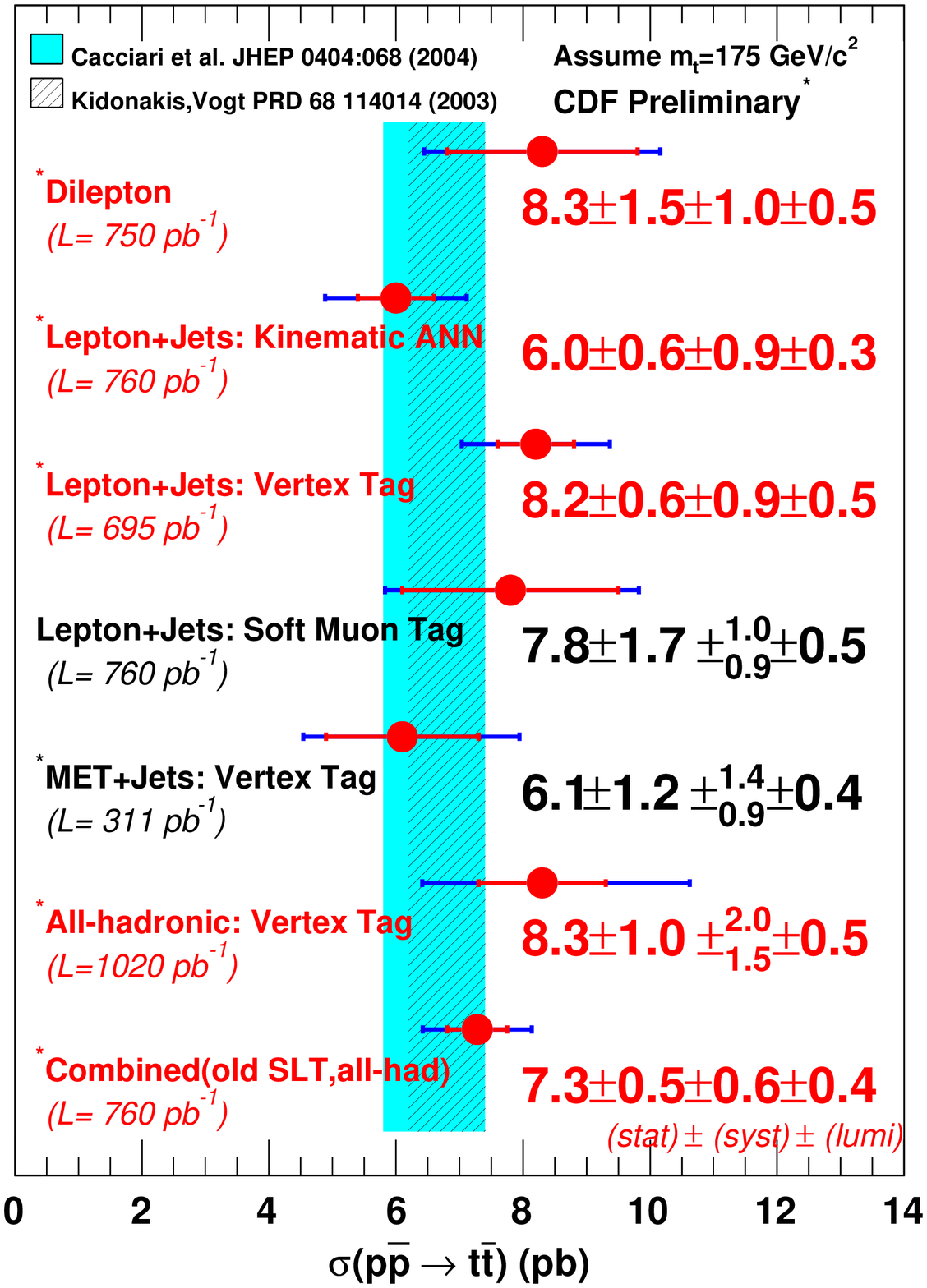,height=3.4in} &
\epsfig{file=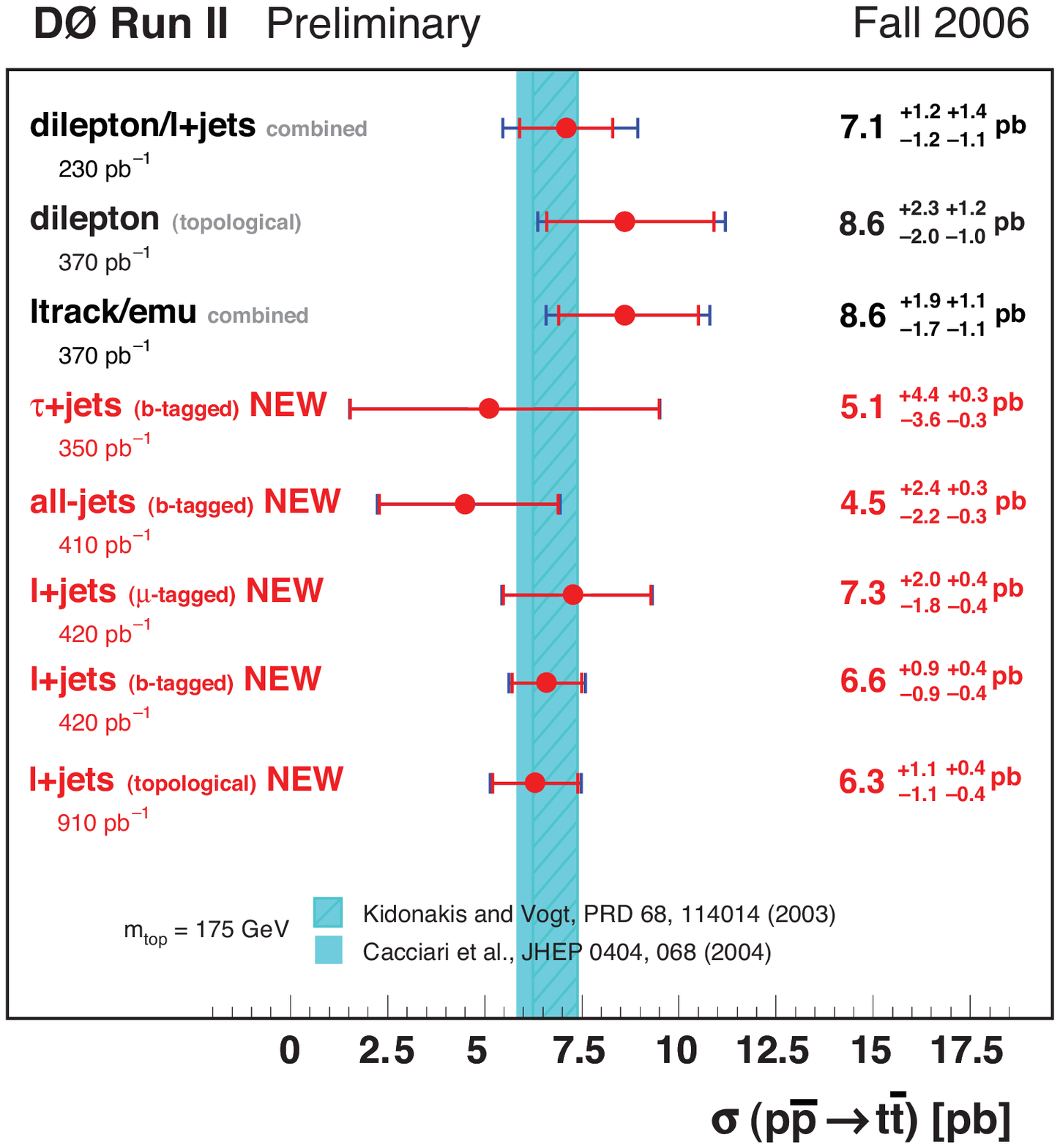,height=3.4in} \\
(a) & (b) \\
\end{tabular}
\caption{Preliminary measurements of the $t\bar{t}$ production cross section at $\sqrt{s}$=1.96 TeV, performed by (a): CDF and (b): D$\oslash$ experiments.}
\label{fig:xsec}
\end{center}
\end{figure}

\subsection{Overview of $t\bar{t}$ production cross section measurements}
The overview of preliminary measurements of the $t\bar{t}$ production cross section at the Tevatron is shown in Figure~\ref{fig:xsec} for the CDF experiment (a) and for the D$\oslash$ experiment (b). The measurements analyse datasets corresponding to integrated luminosity between $\sim$200 pb$^{-1}$ and $\sim$1 fb$^{-1}$. A systematic uncertainty of about 6\%, on the knowledge of the integrated luminosity of the samples, is common to all measurements for each experiment. All the results are in good agreement with the SM calculation, and the measured cross section averaged over different decay channels has now reached a similar accuracy ($\sim$12\%) to the theoretical prediction. Moreover, the measured cross section appears, within the current uncertainties, in good agreement among the different decay channels.

\section{Top Quark Pole Mass}
\subsection{Dilepton events}\label{sec:3.1}
The dilepton channel, consisting of the decays $t\bar{t}\rightarrow\bar{b}\ell^-\bar{\nu}_\ell b\ell^{\prime +}\nu^\prime_\ell$, has a small branching fraction but allows measurements which are less reliant on the calibration of the jet energies than measurements in channels with hadronic $W$ decays. This analysis is based on an integrated luminosity of $\sim$1.0 fb$^{-1}$ collected with the CDF II detector between March 2002 and March 2006. Top candidate events are selected by requiring two leptons, both with $E_T>$ 20 GeV ($P_T>$ 20 GeV/$c$ form muons) and at least one of which is isolated (\footnote{An isolated lepton is one for which no more than 10\% extra energy is measured in a cone of $\Delta R\equiv\sqrt{(\Delta\phi)^2+(\Delta\eta)^2}\le0.4$ around the lepton}). Candidate events must also have at least two jets with $E_T>15$ GeV, measured within $|\eta|<$2.5. We also require candidate events to have $/\!\!\!\!E_T>25$ GeV and, in events with $/\!\!\!\!E_T<50$ GeV, that the $/\!\!\!\!E_T$ vector is at least 20$^\circ$ from the closest lepton or jet. In order to extract maximum information from the sample, we adapt in this analysis a technique based on leading-order production cross section and a parametrized description of the jet energy resolution. Per-event likelihoods in top mass are combined to construct a joint likelihood from which the top quark mass ($M_t$) is determined. The total expression for the probability of a given pole mass for a specific event can be written as:
\begin{equation}
P(x|M_t)=\frac{1}{N}\int d\Phi_8|{\cal M}_{t\bar{t}}(p;M_t)|^2\prod_{\rm jets} W(p,j)W(p_T,U) f_{\rm PDF}(q_1) f_{\rm PDF}(q_2),
\label{equ:ME}
\end{equation}
where the integral is over the entire six-particle phase space, $q$ is the vector of incominig parton-level quantities, $p$ is the vector of resulting parton-level quantities: lepton and quark momenta. The functions $W$ are called transfer functions, and parametrise the probability of measuring a detector-level observable ($j$, $U$) given a parton-level observable ($p$, $p_T$). Finally, $|{\cal M}_{t\bar{t}}(p;M_t)|$ is the $t\bar{t}$ production matrix element as defined in \cite{ttbarnlo,ttbarnlo2}. Background events are taken into account by constructing a generalized per-event probability which includes terms for each background, calculated with a differential cross section similar to Equation \ref{equ:ME}. After calibrating the technique with Monte Carlo simulations of the experiment, we measure: $M_{\rm top}$=164.5$\pm$3.9(stat)$\pm$3.9(syst) GeV/$c^2$. The single largest source of systematic uncertainty comes from the uncertainty in the jet energy calibration ($\pm$3.5 GeV/$c^2$). Other significant systematic uncertainties are due to the Monte Carlo modeling of $t\bar{t}$ production and decay, and the modeling of the background events.

\subsection{Lepton plus jets events}
This analysis is a preliminary measurement of the top quark mass from the lepton plus jets decay channel using $\sim$940 pb$^{-1}$ of data collected from March 2002 to February 2006 with the CDF II detector. Events are selected requiring a single, high-transverse energy (greater than 20 GeV), well-isolated lepton, large missing energy from the neutrino (greater than 20 GeV), and exactly four central jets with high transverse energy (greater than 15 GeV, $|\eta|<2.0$), two from the $b$ quarks and two from the hadronic $W$. Of these jets we require at least one to be identified as originating from a vertex displaced from the primary vertex. In order to reject most non-$W$ background, we further require the axis between missing transverse energy and leading jet, $\Delta\phi$, not be collinear for the lowest values of missing transverse energy passing our selection. The method to determine the top quark mass is similar to the one described in Section \ref{sec:3.1} for dilepton events, whereby we create a likelihood for each event by combining a signal probability with a background probability. However, in this analysis the combined likelihood describes, and is maximised for, not only the top quark mass but also the jet energy scale (JES, an overall multiplicative factor to the jet energy measurements), and the fraction of events consistent with the signal hypothesis. The JES systematic uncertainty is measured constraining the mass of the $W$ from the two untagged jets to 80.4 GeV/$c^2$, and we assume the JES determined for the W-jets also applies to $b$-jets. An additional systematic uncertainty is included to take into account the difference between the JES determined from the hadronic decaying $W$, and the proper $b$-jets energy scale. Simulated samples with different top quark masses as input are used to validate and calibrate the anaysis method. The extracted top mass is found to be unbiased with respect to the top quark mass for which the input events were generated. From 166 events passing our selection requirements we measure: $M_{\rm top}$=170.9$\pm$2.2(stat+JES)$\pm$1.4(syst) GeV/$c^2$. The quoted systematic error include uncertainties on the Monte Carlo modeling of $t\bar{t}$ production and decay, the detector response, and the modeling of background events.

\subsection{All hadronic events}
This analysis is performed using $\sim1$ fb$^{-1}$ of $p\bar{p}$ collisions at $\sqrt{s}=1.96$ TeV collected with the CDF II detector. The technique used is based on a comparison, through likelihood maximization, of the top invariant mass distribution in data with a Monte Carlo simulation of the top signal and background events. The event selection is described in Section \ref{sec:2.2}. We define a quantity ($\chi^2$) as a function of one free parameter, the top mass itself, and use the first 6 jets in order of decreasing $E_T$ to fully reconstruct the event. Of the possible permutations of jet assignments compatible with the $t\bar{t}$ decay chain, we choose the one with the lowest $\chi^2$. The $\chi^2$ contains two terms which constrain the light quark jets to form the two $W$ masses. Then a third jet four-momentum is added in order to form two objects closely spaced in the unknown mass. An additional term is added to account for the uncertainties on jet momentum measurements. The definition is thus as follows:
\begin{eqnarray}
\chi^2 & = & \frac{(m_{jj_1}-m_W)^2}{\Gamma_W^2}+\frac{(m_{jj_2}-m_W)^2}{\Gamma_W^2}+\frac{(m_{jjj_1}-m_t)^2}{\Gamma_t^2} \nonumber \\
& & +\frac{(m_{jjj_2}-m_t)^2}{\Gamma_t^2}+\sum_{i=1}^{N}{\frac{(p_i^{\rm fit}-p_i^{\rm data})^2}{\sigma_i^2}}, \nonumber \\
\end{eqnarray}
where $m_{jj}$ is the invariant mass of the dijet, $m_{jjj}$ is the invariant mass of the trijet, $m_W=80.4$ GeV/$c^2$ and $\Gamma_W$=2.1 GeV/$c^2$ the $W$ boson mass and width, $\Gamma_t$=1.5 GeV/$c^2$ and $m_t$ the free parameter, with the constraint that the mass of the two top quarks be equal. The distribution of invariant mass for background events is derived from the data using the sample before the $b$-tagging request, and is then corrected for the presence of top events. Given a model for signal and background mass templates, we use a likelihood function to determine the mass of the top quark that best describes the data, as well as extracting the number of signal ($n_s$) and background ($n_b$) tags in the sample. We measure $n_s=334\pm33$ and $n_b=573\pm26$ tags and the top quark mass of $M_t$=174.0$\pm$2.2(stat)$\pm$4.8(syst) GeV/$c^2$. Systematic uncertiainties are largely due to the uncertainty on the jet energy measurements ($\pm$4.5 GeV/$c^2$), with smaller contributions ($\le$1.0 GeV/$c^2$) from the Monte Carlo modeling of the top signal and the detector response.

\subsection{Overview and combination of the top quark mass measurements}
\begin{figure}[htb]
\begin{center}
\epsfig{file=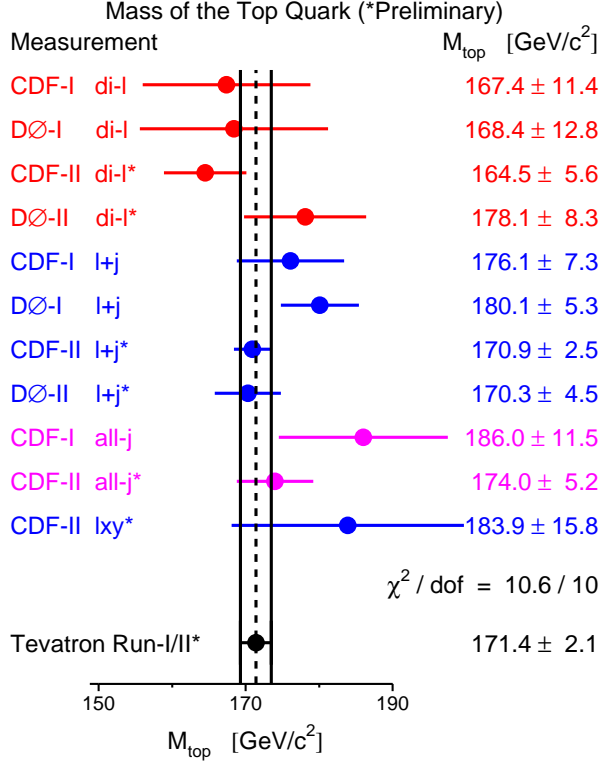,height=4in} 
\caption{Summary of the measurements included in the world average mass of the top quark.}
\label{fig:mass}
\end{center}
\end{figure}

The recent preliminary measurements of the top quark mass at CDF, together with the measurements of the D$\oslash$ experiment, are summarized in Figure \ref{fig:mass}\cite{tevaverage}. The plot includes published measurements from Tevatron data in Run I (1992-1996) and preliminary measurments in Run II, in all three decay channels of the top quark pair. The measurements in the different channels are reasonably consistent with each other. The combination of the measurements, taking into account the statistical and systematical uncertainties and their correlations, yields the preliminary world average mass of the top quark: $M_t$=171.4$\pm$1.2(stat)$\pm$1.8(syst) GeV/$c^2$, which corresponds to a total uncertainty of 2.1 GeV/$c^2$. The precision on the mass measurement is limited by the systematic uncertainties, which are dominated by the calibration of the jet energy measurements.

\section{Conclusions}
We have presented recent preliminary measurements of the top quark pair production cross section and the top quark mass at the Tevatron Collider. All recent measurements are now using $\sim$1 fb$^{-1}$ of proton-antiproton collision data at a center-of-mass energy of $\sqrt{s}=1.96$ TeV. All $t\bar{t}$ cross section measurements are in good agreement with the SM calculation, and the precision of the CDF average cross section is approaching the theoretical accuracy of $\sim\pm$12\%. The world average mass of the top quark is: $M_t$=171.4$\pm$1.2(stat)$\pm$1.8(syst) GeV/$c^2$ and is therefore now known with an accuracy of 1.2\%. The systematic uncertainty, which limits the precision to the mass measurement, is expected to improve as larger data sets are collected at the Tevatron Collider. It can be reasonably expected that with the full Run II data set the top quark mass could be known to better than 1\%.


\end{document}

%% file: econfmacros.tex
\def\beq{\begin{equation}}
\def\eeq#1{\label{#1}\end{equation}}
\def\eeqn{\end{equation}}

\def\beqa{\begin{eqnarray}}
\def\eeqa#1{\label{#1}\end{eqnarray}}
\def\eeqan{\end{eqnarray}}

\let\bar=\overbar

\def\Dslash{\not{\hbox{\kern-4pt $D$}}}
\def\dslash{\not{\hbox{\kern-2pt $\del$}}}

\def\msb{{\bar{\ssstyle M \kern -1pt S}}}




%% file: hql06-lcerrito.bbl
\begin{thebibliography}{99}
\bibitem{definitions}
The transverse momentum, $P_T$, is the projection of a particle's momentum onto the plane perpendicular to the beam axis. The pseudorapidity, $\eta$, is defined as: $-{\rm ln}~{\rm tan}(\theta/2)$, where $\theta$ is the polar angle with respect to the direction of the proton beam. The missing transverse energy, $/\!\!\!\!E_T$, is the transverse component of the energy imbalance in the calorimeter, and is identified with the undetected neutrino's transverse momenta.
\end{thebibliography}
